\documentclass[12pt]{article}
\usepackage{natbib}
\usepackage{graphicx}
\usepackage[table]{xcolor}
\usepackage{float}
\usepackage{xcolor}
\usepackage{url}
\usepackage{bm}
\usepackage{tikz} 
\usepackage{pifont} 
\usepackage{amsfonts}
\usepackage{amssymb}
\usepackage{amsmath,upgreek}
\usepackage{titlesec}
\usepackage{url} 
\usepackage{longtable}
\usepackage{multirow}
\usepackage{fixfoot}
\usepackage{fixfoot,xspace}
\usepackage{lscape}
\usepackage{blindtext} 
\usepackage{booktabs}
\usepackage{authblk}
\usepackage{caption}

\title{Binary asteroids in mean-motion resonances}
\author[1]{Nata\v sa Todorovi\' c}
\author[1]{Ivana Mili\' c \v Zitnik}

\affil[1]{Astronomical Observatory, Volgina 7, 11060 Belgrade 38, Serbia}
\date{}

\begin{document}
\maketitle

\begin{abstract}
The purpose of this study is to investigate the relation between binary asteroids 
and mean motion resonances (MMRs). For more than 700 asteroids from two catalogues, the \textit{Johnston Archive} \citep{johnstons} and the \textit{Gaia DR3 VizieR} list of binary  candidates from \cite{Liberato2024}, we applied a resonance identification algorithm, treating 
all planetary perturbations. Our results showed that the presence of binary asteroids in MMRs 
largely depends on their dynamical class. The highest percentage, more than 30\%, is found in the 
Trans-Neptunian region where most of these objects have exhibited 
resonant librations longer than 10~Myr. For the main-belt asteroid pairs, this percentage is 
about 10-12\%. Contrary to expectations, the more unstable region populated with  NEOs, 
showed a higher percentage of resonant pairs (above 17\%), but with temporal resonant 
captures. These results could indicate that the mean motion resonances, particularly the stronger 
ones, could play a role in the evolution and formation  of binary systems. Finally, we highlight 
that in the present paper, 82 resonant binary asteroids are newly identified. 
\end{abstract}

\footnotetext[0]{\copyright \ 2025 The Author(s). Published in the Serbian Astronomical Journal, by Astronomical Observatory of Belgrade and Faculty of Mathematics, University of Belgrade. This open access article is distributed under CC BY-NC-ND 4.0 International licence.}

\section{Introduction}

The concept of a binary asteroid was first introduced in 1901 by the French astronomer 
Charles Andr\' e \citep{Andre901, Radau1901BuAs}. According to a NASA/ADS search, the 
first mention of a binary system appeared in \cite{Cook1971}, where the author proposed that 
the unusual light curve of the Trojan asteroid (624) Hector indicated the presence of a 
satellite, rather than an elongated, cigar-shaped body\footnote{Further studies showed that 
both assumptions were correct.} \citep{DunlapGehrels1969AJ}. Over the following years, 
additional asteroids with similar photometric characteristics were identified (see, e.g.~\citeauthor{Binzel1979Sci} 
\citeyear{Binzel1979Sci}, 
\citeauthor{Binzel1985Icar} \citeyear{Binzel1985Icar}, 
\citeauthor{Tedesco1979Sci} \citeyear{Tedesco1979Sci}). However, the existence of binary 
asteroids remained speculative 
(\citeauthor{Binzel1978} \citeyear{Binzel1978}, 
\citeauthor{vanFlandern1979} \citeyear{vanFlandern1979}, 
\citeauthor{Weidenschilling1989} \citeyear {Weidenschilling1989})
until 1993, when the Galileo spacecraft directly imaged the (243) Ida-Dactyl asteroid pair 
\citep{Mason1994, Belton1996Icar}.

The classical method for the identification of binary and multiple systems\footnote{In the 
following text, for simplicity, we will use only the term \textit{binaries}.} is by tracking 
signatures in their light curves, and it applies to all dynamical classes (although not to widely 
separated pairs). Faint and distant main belt and trans-Neptunian binaries require more 
advanced observational methods and are also found by direct imaging or stellar occultations. 
Near-Earth binaries are often identified on radar images, whereas a small number of systems 
have been imaged directly from spacecraft. Additional detection methods rely on spectral 
analysis, astrometric observation residuals,  similarities in heliocentric orbits, or analysis of  
rotational properties \citep{Noll2023, Margot2015, MarcosMarcos2019}. 

Binary systems exhibit significant diversity, to the extent that nearly every pair could be 
studied separately. However, certain characteristics, particularly those related to formation, 
are often shared within the same dynamical classes.  
Binary Trans-Neptunian Objects (TNOs) are likely to have a primordial origin. Numerous 
studies 
\citep{Weidenschilling2002, Goldreich2002Natur, NYR2010AJ, Fraser2017NatAs, 
Robinson2020AA, Nesvorny2021PSJ} investigated models in which the gravitational collapse of 
pebble clouds transferred the angular momentum to the formation of binary planetesimals. 
Being less exposed to major planetary perturbations which does not exclude mutual 
gravitational interactions 
\citep{Brunini2020, Lopez2021MNRAS, Lawler2024}, these systems 
could have survived to the present day, accounting for the high fraction of observed binary 
TNOs \citep{Noll2008book, Lawler2024}. 

In regions closer to the Sun, different scenarios unfold. Main-belt binaries are more likely to 
be formed in post-collision processes, either by reaccumulation of ejected material or mutual 
gravitational capture of their larger fragments \citep{Michel2001Sci, Durda2004Icar, 
Durda1996Icar, Doressoundiram1997, Giblin1998Icar}. Small Near-Earth asteroids (NEAs) 
affected by the YORP (Yarkovsky-O'Keefe-Radzievskii-Paddack) thermal force 
\citep{Rubincam2000, VokrouCapek2002Icar, Bottke2006} are largely studied in the context of 
binary formation (see, e.g.~\citeauthor{Cuk2007ApJ} \citeyear{Cuk2007ApJ}, 
\citeauthor{PravecHarris2007} \citeyear{PravecHarris2007}, 
\citeauthor{Scheeres2007Icar} \citeyear{Scheeres2007Icar}, 
\citeauthor{WalshRichardson2008} \citeyear{WalshRichardson2008}, 
\citeauthor{WalshRM2008Nature} \citeyear{WalshRM2008Nature}, 
\citeauthor{JacobsonScheeres2011Icar} \citeyear{JacobsonScheeres2011Icar}, 
\citeauthor{Wimarsson2024Icar} \citeyear{Wimarsson2024Icar}). 
Under the action of YORP, the primary asteroid spins up to a rate high enough to shed 
mass, which then coalesces into a secondary body, forming a binary system\footnote{This 
scenario refers to rubble pile asteroids.}. Reported formation times vary widely, from $10^5$ 
years to only a few hours \citep{Wimarsson2024Icar}. Asteroid pairs can form even 
'instantaneously', in scenarios where a larger piece falls off from a rubble-pile asteroid due to 
rotational fission \citep{Scheeres2009CeMDA, JacobsonScheeres2011Icar}.

Alternative formation mechanisms include tidal forces or migrations from the main asteroid 
belt  \citep{WalshRichar2006Icar, WalshRichardson2008}, but it is unlikely that 
primordial pairs exist in the dynamically unstable NEA region. However, double craters on 
terrestrial planets and the Moon \citep{MeloshStansberryIcar1991, Cook2003Icar, 
Miljkovic2013EPSL, Vavilov2022} indicate their larger presence in the past. 

According to our knowledge, no systematic search of binary asteroids in mean motion 
resonances (MMRs) has been done so far. Previous works treated only individual cases; for 
example, \cite{Rosaev2024CeMDA} investigated resonant perturbation of the pair (5026) 
Martes and 2005 WW113 caused by the 3E-11 MMR with Earth.  \cite{Borisov2024CoSka} 
captured a binary candidate 12499 exactly on the chaotic border of the 4M-7 MMR with Mars. 
Also, \cite{Pravec2019Icar} found irregular jumps of the pair (10123) Fide\" oja - 117306, over 
the 7J-2 MMR, and possible instability of the pair 49791 - 436459 due to the 8M-15 MMR. 
\cite{Pravec2019Icar} also discussed pairs in secular and spin-orbit resonances, but they are 
not the subject of this research. \cite{Duddy2012AA} identified the unknown resonance 
interacting with the pair 7343-154634 \citep{PravecVokrouhlicky2009Icar}, to be the  1M+1J-2 
three-body MMR between the asteroid, Jupiter and Mars. 

The search for binaries among the resonant Trojan (1J-1) and Hilda (3J-2) populations in the 
NEOWISE archive was performed by \cite{Sonnett2015ApJ}. The authors found a surprisingly 
high binary fraction, of 14\%–23\% among the Trojan asteroids larger than $\sim 12$ km and 
even $30\%-51\%$ among Hildas larger than $\sim 4$ km.  However, these results are not 
entirely reliable, as only the binary \textit{candidates}, i.e., asteroids with large light curve 
amplitudes, were considered.

In the Kuiper belt, \cite{Compere2013AA} investigated the scarcity of binaries in the Plutino 
population i.e. in the 2N-3 resonance with Neptune, although newer studies report a high 
fraction of contact and tight Plutino binaries \citep{Thirouin2018AJ, Brunini2023MNRAS}; 
also \cite{Thirouin2024PSJ} found several contact binaries in the 3N-5 and 4N-7 MMRs. 

Resonances, particularly the stronger ones, may have played a role in the evolution and 
possibly the formation of binary systems. Dynamical description of these processes is a 
challenging task for future work. Here, we aim to make the first step: find all known asteroid 
pairs captured by mean motion resonances and see whether their presence is larger in MMRs.

\section{Methodology}
\label{methodology}

\indent

We use two catalogues of binary asteroids: the Johnston 
Archive\footnote{https://www.johnstonsarchive.net/astro/asteroidmoons.html} 
\citep{johnstons}, with 477 objects  (in 
October 2024) spanning from the Near Earth to the trans-Neptunian region, and a recently 
published list from Gaia DR3 dataset \citep{Liberato2024} containing  357 astrometric binary 
candidates, mostly from the main belt. All asteroids were integrated with the astronomical 
package \texttt{resonances} \citep{Smirnov2023} for $100$ Kyr, which is the default time of 
the software. A certain number of trans-Neptunian objects was additionally integrated with 
Orbit9\footnote{http://adams.dm.unipi.it/orbfit/} for 10 Myr. Perturbations from all planets 
were included in all calculations.

Following the algorithm suggested by \citet{SmirnovDovPop2017}, we calculated all potential 
two and three-body mean-motion resonances for each asteroid, using the proximity between 
the asteroid's and resonant semi-major axes. For every asteroid-MMR pair, we performed an 
identification procedure to determine whether the asteroid was trapped in the given 
resonance. The identification procedure requires two main conditions to be met: 

\begin{itemize}
\item [\textit{i)}] The resonant argument $\sigma$ must librate during the 
integration. For two-body resonances $\sigma$ is defined with $\sigma = m_p \lambda_p + 
m\lambda + n_p\varpi_p +n\varpi$, where $\lambda_p$ and $\lambda$ are mean longitudes 
of the planet and the asteroid; $\varpi_p$ and $\varpi$ are the longitudes of their pericenters, 
and the integers $m_p$, $m$, $n_p$, and $n$ satisfy the d’Alembert criterion. For three-body 
resonances, appearing with commensurability between an asteroid and two planets, $\sigma$ 
is defined with $\sigma = m_{p_1}\lambda_{p_1} +  m_{p_2}\lambda_{p_2} + m\lambda + 
n_{p_1}\varpi_{p_1}+n_{p_2}\varpi_{p_2}+n\varpi $. The variables are the same as for two-body 
resonances, with the difference that indices $p_1$ and $p_2$ refer to planets 1 and 2 involved 
in the resonance. We considered only the planar case and the primary subresonance in the 
multiplet, which implies $n_p=n_{p_1}=n_{p_2}=0$.
\item [\textit{ii)}] The oscillation frequencies of the resonant argument and semi-major axis 
should match and the corresponding amplitudes should be sufficiently large to be 
meaningful \citep{Gallardo2006a}.
\end{itemize}

The identification procedure performed by the \texttt{resonances} package is specified in 
several papers: the resonant axis calculation and classifications of librations for three-body 
MMRS with Jupiter and Saturn, and two-body MMRs with Jupiter in 
\citet{SmirnovShevchenko2013Icar}, the expansion to all possible combinations of planets in 
\citet{SmirnovDovPop2017}, and the Python package used in the 
paper for the identification in \citet{Smirnov2023}. 

\begin{figure*}
\centering
\includegraphics[width=0.45\textwidth]{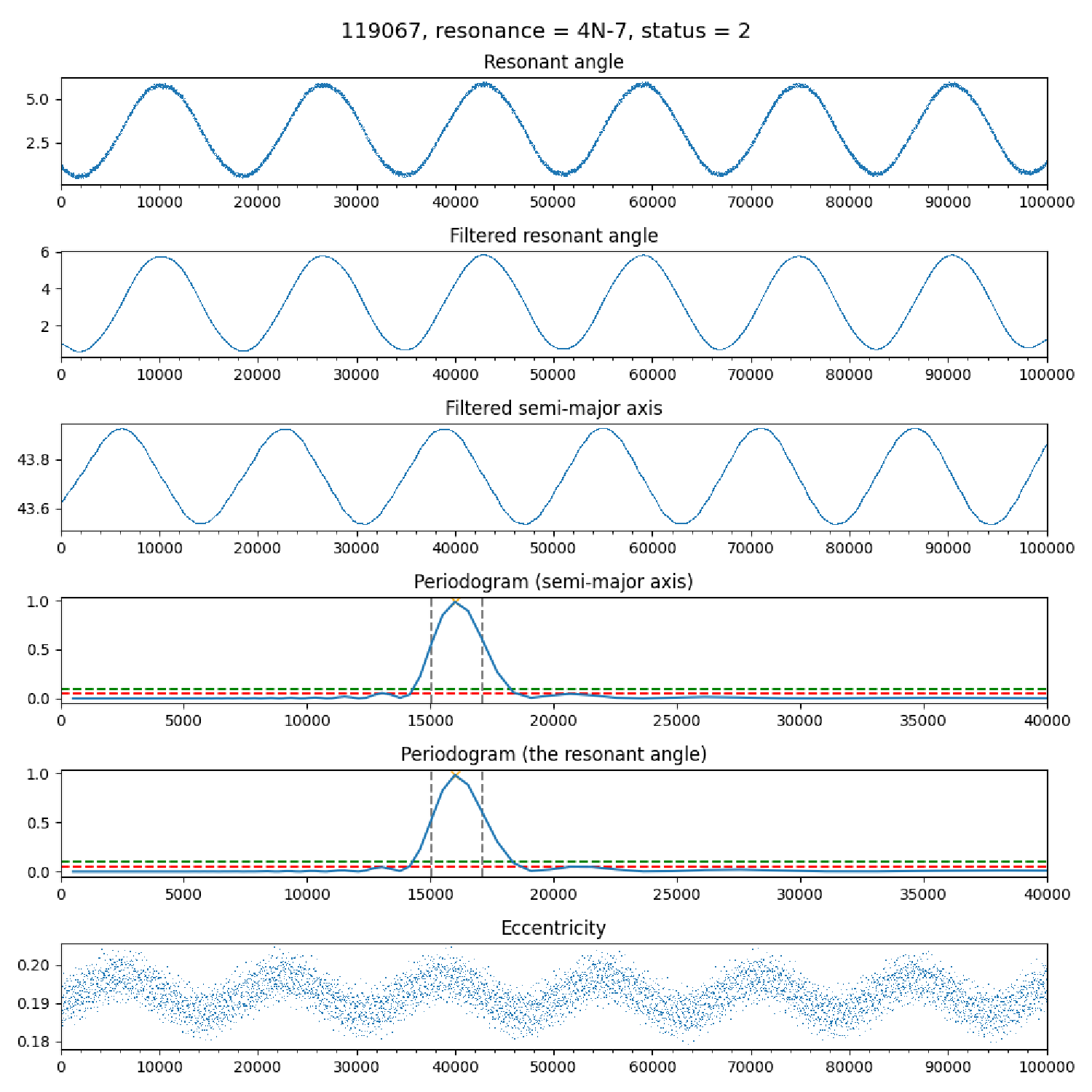}
\includegraphics[width=0.45\textwidth]{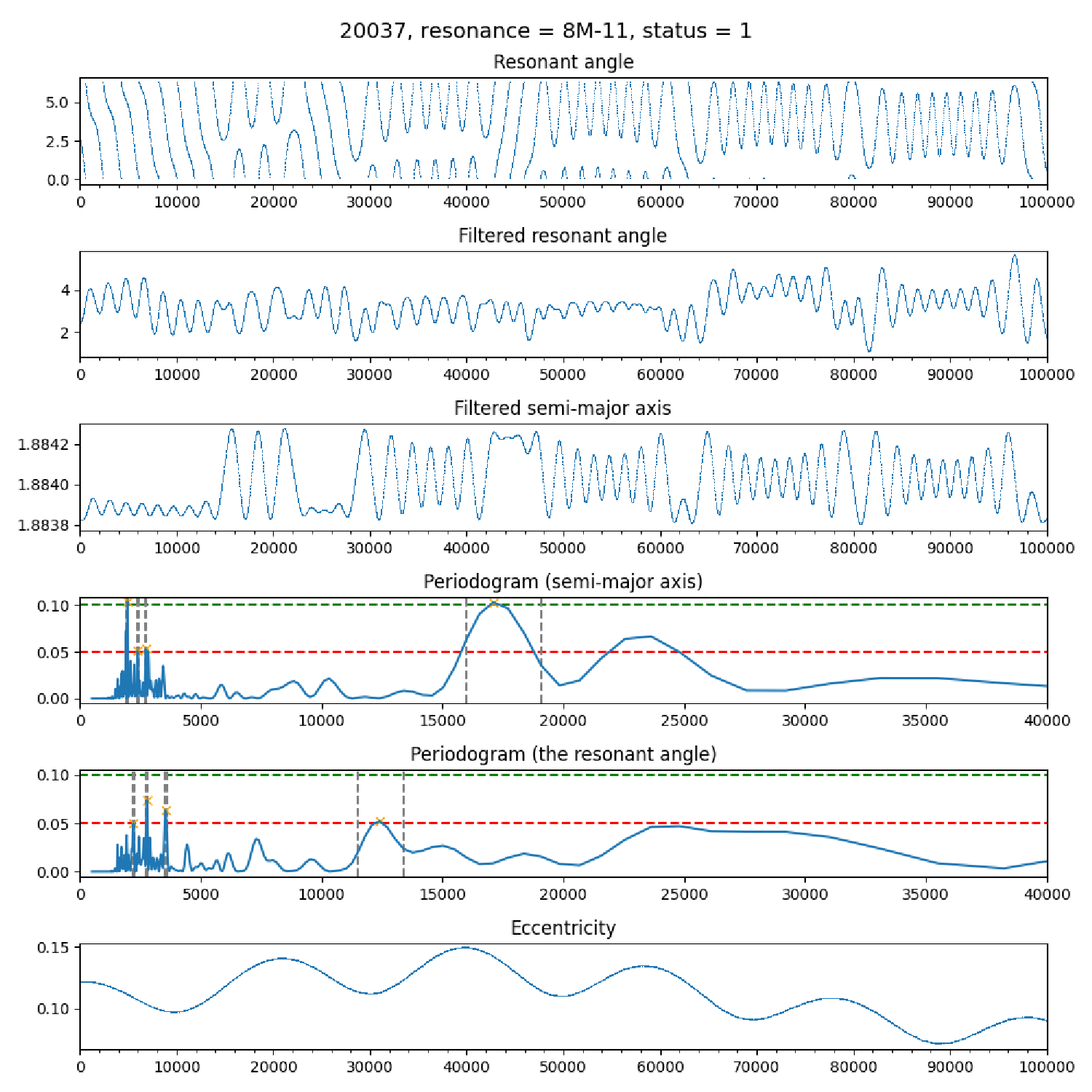}

\caption{\small The left plot illustrates the case of pure \textit{libration} of the asteroid 119067 in the 4N-7 MMR with Neptune. 
The first two rows show that the resonant angle $\sigma$ and its filtered value $\sigma_f$ librate during the whole interval of 100~Kyr, while $a$ oscillates with the same frequency (row 3). The two periodograms (rows 4 and 5) indicate that the dominant period for both $\sigma$ and $a$ aligns at approximately 16,000 years. The last row displays the change in eccentricity with approximately the same period. The right panel shows the so-called \textit{transition} status for the asteroid 20037 in the 8M-11 MMR with Mars. The asteroid is trapped in the resonance in several episodes between  $ t= \{[14,24], [28,42], [48,60], [63,98] \}$ Kyr when $\sigma$ and $\sigma_f $ (rows 1 and 2) librate, and $a$ oscillates with somewhat larger amplitudes (row 3). The two periodograms (rows 4 and 5) show a dense period distribution for times less than 5000 years, with the most dominant peaks at 2 kyr, 2.5 Kyr, and 3 Kyr. The eccentricity of the asteroid (row 6) illustrates a periodic change, but also a slight average increase in the first 50~Kyr.}
\label{fig1}
\vspace{-2mm}
\end{figure*}

After completing the integration, the software automatically detects the resonant and potentially 
resonant asteroids and generates corresponding images for each asteroid (see Fig.~\ref{fig1}). 
The images are composed of six panels showing the time evolution of resonant angle 
$\sigma$ (row 1), its filtered value $\sigma_f$ (row 2) where all short periodic oscillations with periods less 
than 500 years are removed, semi-major axis $a$ [AU] (row 3), and eccentricity (row 6). The 
software also presents two periodograms that illustrate the distribution of all periodic 
variations in $a$ and $\sigma$ (rows 4 and 5), where the red and green dashed horizontal 
lines represent the weak and strong thresholds for periods to be significant. For a more 
detailed explanation of periodograms and filtering procedures, see 
\citet{SmirnovDovgalev2018} and the documentation of the \texttt{resonances}\footnote{Available at: 
https://smirik.github.io/resonances/libration/} software. The images produced by the 
software were used by the 
authors for visual inspection for additional confirmation that the asteroids are in MMRs. 
Positive cases are classified into two categories:

\begin{itemize}
\item [a)] \textit{Libration}: the asteroid librates in the resonance throughout the 
\textit{entire} integration period. 
\item [b)] \textit{Transition}: the asteroid is temporarily captured in the resonance and 
librates in \textit{episodes}. 
\end{itemize}

\vskip-1mm

\textit{Libration} of the asteroid 119067 in the 4N-7 MMR with Neptune is illustrated in the 
left panel of Fig.~\ref{fig1}. The top two rows show that both $\sigma$ and $\sigma_f$ 
librate during all 100 Kyr, while the semi-major axis oscillates around $a \sim 43.7$ [AU] with 
approximately the same frequency (row 3). Periodograms for $a$ and $\sigma$ (rows 4 and 5) 
give an additional confirmation of resonant status, since they have only one dominant 
peak at about 16 Kyr. Row 6 gives the asteroid's eccentricity time change, with approximately 
the same period. Pure librations are marked by the software with 2, as noted at the top of the 
panel.

An illustration of the \textit{transition} status of the asteroid 20037 in the 8M-11 MMR is given 
in the right panel of Fig.~\ref{fig1}. The first three rows $\sigma$, $\sigma_f$ and $a$ 
indicate that the asteroid entered and left the resonance several times. Librations in $\sigma$ 
are observed between [14,24] Kyr, [28,42] Kyr, [48,60] Kyr and [63,98] Kyr, during which $a$ 
oscillates with larger amplitudes.  The distribution of oscillation periods in rows 4 and 5 
shows a higher concentration for $t < 5 $ Kyr, with three matching peaks at $ t \sim [2, 2.5, 3]$ 
Kyr. Eccentricity displays a periodic change, but also a slight increasing trend. The software 
marked the transient resonant status with 1, which is displayed at the top of the panel.

This study is mainly intended to detect asteroids that are already affected by the resonance, 
meaning that 100,000 years  (the default time of the software) can cover the necessary interval 
for identifying resonant behaviours. For the TNO population, we have repeated the 
integration of purely resonant objects for 10 Myr, using Orbit9 and the same parameters as 
used by the \texttt{resonances} package.

\section{Results and discusssion}
\label{results}

\indent

Table~\ref{table1} provides a summary of the results, presenting the binary asteroids from both 
the Johnston Archive (hereafter JA) and the Gaia catalogue, categorised by their dynamical 
classes. It shows the presence of resonant objects within each group, as well as within the 
overall JA and Gaia binary sets. Detailed lists of resonant binaries, their associated resonances, 
resonant locations and assigned statuses are given in Table~\ref {table2} for JA, and 
Table~\ref{table3} for the Gaia set. 

\begin{table*}[t!]

{\ }
\caption{This table summarises the results on the resonant binaries search from the Johnston 
Archive (JA) and the Gaia DR3 dataset, obtained with the \texttt{resonances} package over the 100Kyr 
integrations. Dynamical types and the number of objects in them are in column 2 and column 3, 
respectively. The number of resonant objects in each class is given in column 4, 
while column 5 gives the number of objects in libration (L) and transient (T) regimes. Column 6 gives the percentage of resonant objects in each of these groups.}

\vspace{.4cm}
\small
\centerline{\begin{tabular}{l|c|c|c|cc|c}
\hline
  \arrayrulecolor{gray} \hline
\rowcolor{brown!10} Sourse  & Dynamical type & No. of objects & No. of objects in MMRs & Res.  status &  & Result in [\%] \\
    \hline
    \hline    
\multirow{2}{*}{JA}  &\multirow{2}{*}{NEA}          & \multirow{2}{*}{82} &  \multirow{2}{*}{14}  &  L     & 1   &  \multirow{2}{*}{17.1} \\
                     &                              &                     &                       &  T     &  13 &                        \\
 \hline 

\multirow{2}{*}{JA}   & \multirow{2}{*}{Mars-cros.} & \multirow{2}{*}{34}  &\multirow{2}{*}{6}    & L & 2 & \multirow{2}{*}{17.6} \\ 
                                 &                   &                     &                      & T &4 &    \\ 
              
  \hline     
\multirow{2}{*}{JA}   & \multirow{2}{*}{Main belt}   & \multirow{2}{*}{268}  & \multirow{2}{*}{29}   & L  & 3 &  \multirow{2}{*}{10.8} \\ 
                          &                              &                   &                     & T & 26 &   \\

  \hline 
\multirow{2}{*}{JA}   & \multirow{2}{*}{Trojans}     & \multirow{2}{*}{8}    & \multirow{2}{*}{8}    & L & 8  &  \multirow{2}{*}{100} \\
                                 &                       &                   &                      & T & 0  & \\

 \hline       
\multirow{2}{*}{JA}   & \multirow{2}{*}{TNOs}        & \multirow{2}{*}{85}   & \multirow{2}{*}{26}   & L & 25  & \multirow{2}{*}{30.5}  \\
                          &                              &                   &                       & T & 1 &    \\

\hline
 \hline        
\multirow{2}{*}{JA}   & \multirow{2}{*}{all}     &\multirow{2}{*}{477}       &\multirow{2}{*}{83}   &  L& 39 &  \multirow{2}{*}{17.4}  \\
  &                              &                   &                      &  T  &  44 &    \\
 \hline         
 \hline     

 \hline    
\multirow{2}{*}{Gaia}   & \multirow{2}{*}{NEA}     & \multirow{2}{*}{2}    & \multirow{2}{*}{0}    & L& - &  \multirow{2}{*}{-} \\
                        &                              &                   &                       & T & - &   \\
\hline
\multirow{2}{*}{Gaia}   & \multirow{2}{*}{Mars-cros.} & \multirow{2}{*}{8}  &\multirow{2}{*}{0}    & L& - & \multirow{2}{*}{-} \\ 
                                 &                   &                     &                      & T &- &    \\ 
\hline 
\multirow{2}{*}{Gaia}   & \multirow{2}{*}{Main belt}   & \multirow{2}{*}{341}  & \multirow{2}{*}{42}  & L  & 12 &  \multirow{2}{*}{12.3} \\ 
                          &                              &                   &                       & T & 30 &   \\
\hline
\multirow{2}{*}{Gaia}   & \multirow{2}{*}{Trojans}     & \multirow{2}{*}{6}    & \multirow{2}{*}{6}    & L& 6  &  \multirow{2}{*}{100} \\
                                 &                       &                   &                        & T & 0  & \\ 
 \hline     
 \hline   
                                 
 \multirow{2}{*}{Gaia}   & \multirow{2}{*}{all}       &\multirow{2}{*}{357}   &\multirow{2}{*}{48}   & L & 18 &   \multirow{2}{*}{13.4} \\
                        &                          &                   &                      & T  & 30 &   \\
                                 
 \hline     
 \hline     
\end{tabular}}
\label{table1}
\end{table*}
\vskip3mm

\begin{figure}
\centerline{\includegraphics[width=0.7\columnwidth]{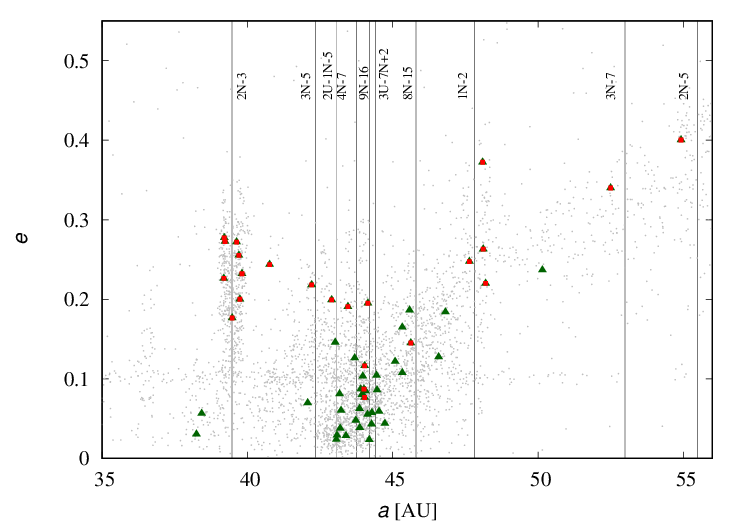}}
\caption{ \small Binary resonant (red triangles), binary non-resonant (green triangles), and 
non-binary numbered (grey dots) asteroids in the TNO region. The vertical dashed lines 
mark the positions of semi-major axes for the most dominant MMRs in the region. The 
largest concentration of resonant binaries is found in the most populated TNO region 
between 44 and 45 AU, and in the most populated 2N-3 MMR in the region.}
\label{fig2}
\end{figure}

Trojan binaries are in the 1J-1 MMR by definition.
All 8 Jupiter Trojans from JA (Table~\ref{table2}) and  6 from Gaia
(Table~\ref{table3}) have pure librations (L).
Besides them, the largest fraction of binaries in MMRs (30.5\%) is found among TNOs. This 
result aligns with earlier studies, which suggest that the primordial origin of TNO binaries 
may be preserved in resonances, shielding these systems from dynamical disruption 
throughout the Solar System's history. The distribution of the binary TNOs in the a-e plane 
between 35 and 56 AU is shown in Fig.~\ref{fig2} (for comparison, see Fig.~1 in 
\citeauthor{Lawler2024} \citeyear{Lawler2024}). Resonant binaries (red triangles) appear to be  prevalent in the densest 
parts of the TNO region, particularly between 44 and 45 AU. The 2N-3 MMR, for which  
\cite{Forgacs-Dajka2023ApJS} found to hold a significant amount of TNOs, also dominates in 
a number of binary resonant objects (in Table~\ref {table2}, 8 TNOs are in the 2N-3 MMR, 
all have pure librations). Let us note that this resonance also captures a significant number of 
contact binaries \citep{Thirouin2018AJ, Brunini2023MNRAS}, which are not contained in  
JA. 

Almost all TNOs (except one - 65489 Ceto) listed in Table~\ref{table2} convincingly reside in 
their resonances exhibiting pure librations during 100 Kyr, suggesting a long-term stay. We 
repeated the calculation of these asteroids for 10 Myr using Orbit9, and found that 19 out of 
25 TNOs remained in pure liberation in the long runs. The six TNOs in Table~\ref{table2} 
designated with 'LT' (column 6) are those objects that have shown pure librations for 100Kyr, 
but passed into transient statuses (T) on the repeated 10 Myr integrations. Each of these 6 
asteroids has actually left and re-entered the resonance multiple times over a period of 10 
million years. A more detailed analysis of this phenomenon and an assessment of stability are 
beyond the scope of this paper and may be examined in a future study.

The main belt pairs in JA and
in the Gaia set are resonant objects
in 10.8\% and 12.3\% cases, respectively,
and most of them in transition statuses.
Even three asteroids in JA (88710, 1803 Zwicky and  78085, see 
Table~\ref{table2}) and three in the Gaia set (43341, 6364 Casarini and 68304, see 
Table~\ref{table3}) resided in more than one resonance during 100 Kyr.
Transitions from one resonance to another most likely arise from resonance overlap in this 
region densely populated with MMRs. In Tables~\ref{table2} and \ref{table3}, we provide the locations of resonances (their centers), but not their widths, which are essential for identifying resonant 
chains that enable such transitions. More studies, which are beyond the scope of this paper, are required to explore these possibilities.

Out of 34 Mars-crossers in JA, 6 (17.6\%) were found in resonances, 1 Mars-Trojan, 5261 
Eureka, and 1 resident of the 2J-1 MMR, the asteroid 8373 Stephengould, exhibit pure 
librations, while the remaining 4 asteroids display a transient nature. The percentage of 
17.6\% is above the main-belt average, but the small sample may render this conclusion 
uncertain.

In the NEA region, almost all resonant pairs (except 452561 in the 2J-1 MMR) have transient 
statuses, showing a high mobility through resonances. It could be expected that these 
dynamical transitions contribute to separation of asteroid pairs, since they are more 
exposed to planetary perturbations and chaos, in general. Surprisingly, the unstable NEA 
region showed a higher proportion of resonant pairs (17.1\%) compared to those in the main 
belt (10.8\% and 12.3\%). This, certainly, is a topic that needs to be investigated. In the Gaia catalogue, 
none of the 2 NEAs or 8 Mars-crossing binaries are found to be in a resonance.

The total percentage of resonant objects in the Johnston Archive is 17.4\%,  undoubtedly 
influenced by the high percentage of resonant TNOs. The mean value for the Gaia set is 
13.4\%. We recall the result from \cite{SmirnovDovPop2017}, who used a similar algorithm for 
resonance identification on all asteroids from the AstDyS database known at that time. They 
found that 14.1\% (65,972 out of 467,303 objects) of asteroids were resonant. However, their 
results were not categorised by dynamical classes, which complicates the direct comparison with the results presented in Table~\ref{table1}. If we compare the overall rates of 17.4\% from JA and 14.1\% from \cite{SmirnovDovPop2017}, since both cover the regions from NEAs to TNOs, we infer that binary asteroids have a greater tendency to reside in MMRs. This could mean that resonances not only protect asteroid pairs but, perhaps, they also \textit{produce} them.  Long stays in stable resonances open up the possibility for mutual encounters, gravitational captures and formation of new pairs. This could explain the high fraction of binary asteroids among Trojans (14\%–23\% larger than 12~km) and Hildas (30\%-51\% larger than 4~km) estimated in \cite{Sonnett2015ApJ}.

\newpage

{\scriptsize
\begin{longtable}{llcllcc}
\captionsetup{width=1.\textwidth}
\caption{\small Binary asteroids from the Johnston Archive identified in various mean motion resonances (listed in column 4). Their dynamical classes, resonant statuses (L for libration or T for transition), and semi-major axis of resonant locations \citep{Gallardo2006a}, are provided in columns 7, 6 and 5, respectively. Column 3 cites the previous studies where these asteroids have been classified as resonant. Out of the 26 TNOs, 6 of them, marked with 'LT', entered into transient statuses in the repeated 10 Myr integrations. Newly identified resonant binary asteroids (44 in total) are indicated with a double asterisk (**).}
\label{table2}
\\
\hline 
\multicolumn{1}{l}{\textbf{No. and name}} & \multicolumn{1}{l}{\textbf{Prov. name }} & 
\multicolumn{1}{c}{\textbf{ Reference}} & \multicolumn{1}{l}{\textbf{MMR}} & 
\multicolumn{1}{c}{\textbf{$a_{MMR}$}} &  \multicolumn{1}{c}{\textbf{status}} &  
\multicolumn{1}{c}{\textbf{class}}\\ \hline 
\endfirsthead 

\multicolumn{7}{l}%
{{Table 2}{: continued from the previous page.}} \\ 
\multicolumn{7}{c}%
{\ }\\
\hline
\multicolumn{1}{l}{\textbf{No. and name}} & \multicolumn{1}{l}{\textbf{Prov. name }} & 
\multicolumn{1}{c}{\textbf{ Reference}} & \multicolumn{1}{l}{\textbf{MMR}} & 
\multicolumn{1}{c}{\textbf{$a_{MMR}$}} &  \multicolumn{1}{c}{\textbf{status}} &  
\multicolumn{1}{c}{\textbf{class}}\\ \hline 
\endhead

\hline \multicolumn{3}{r}{{}} \\ 
\endfoot

\hline 
\endlastfoot

3122 Florence       & (1981 ET3)     &   **                         & 5J-1       & 1.778    &  T   & NEA           \\
 35107               & (1991 VH)      &   **                         & 4E+2J-5    & 1.128    &  T   & NEA           \\
 5646                & (1990 TR)      &   **                         & 3M-5       & 2.141    &  T   & NEA           \\
 7888                & (1993 UC)      &   **                         & 3J-1       & 2.500    &  T   & NEA           \\
 7889                & (1994 LX)      &   **                         & 12E-17     & 1.261    &  T   & NEA           \\
 85804               & (1998 WQ5)     &   **                         & 5M-6       & 1.720    &  T   & NEA           \\
 88710               & (2001 SL9)     &   **                         & 5V-9       & 1.070    &  T   & NEA           \\ 
 88710               & (2001 SL9)     &   **                         & 9V-16      & 1.061    &  T   & NEA           \\ 
 152931              & (2000 EA107)   &   **                         & 7E-6J-6    & 0.948    &  T   & NEA           \\
 163693 Atira        & (2003 CP20)    &   **                         &  4V-15     & 1.745    &  T   & NEA           \\
 164121              & (2003 YT1)     &   **                         & 5M-3S-3    & 1.112    &  T   & NEA           \\
 452561              & (2005 AB)      &   **                         & 2J-1       & 3.276    &   L         & NEA           \\
 458732              & (2011 MD5)     &   **                         & 3J-1       & 2.500    &  T   & NEA           \\
 613286              & (2005 YQ96)    &   **                         & 3E-2       & 1.310    &  T   & NEA           \\
 620082              & (2014 QL433)   &   **                         & 4J-1       & 2.064    &  T   & NEA           \\
 5261 Eureka         & (1990 MB)      &   MPC                        & 1M-1       & 1.523    &   L             & mars crossers \\
 8373 Stephengould   & (1992 AB)      & \cite{RNF2002}               & 2J-1       & 3.276    &   L             & mars crossers \\
 20037 Duke          & (1992 UW4)     &   **                         & 8M-11      & 1.232    &  T   & mars crossers \\
 34706               & (2001 OP83)    &   **                         & 7J-2       & 2.256    &  T   & mars crossers \\
 218144              & (2002 RL66)    &   **                         & 5J-4S-1    & 2.303    &  T   & mars crossers \\
 12008 Kandrup       & (1996 TY9)     &   **                         & 3J+3S-1    & 1.996    &  T   & mars crossers \\
 617 Patroclus       & (A906 UL)      &   MPC                        & 1J-1       & 5.201    &   L             & trojan        \\
 624 Hektor          & (A907 CF)      &   MPC                        & 1J-1       & 5.201    &   L             & trojan        \\
 3548 Eurybates      & (1973 SO)      &   MPC                        & 1J-1       & 5.201    &   L             & trojan        \\
 15094 Polymele      & (1999 WB2)     &   MPC                        & 1J-1       & 5.201    &   L             & trojan        \\
 16974 Iphthime      & (1998 WR21)    &   MPC                        & 1J-1       & 5.201    &   L         & trojan        \\
 17365 Thymbraeus    & (1978 VF11)    &   MPC                        & 1J-1       & 5.201    &   L          & trojan        \\
 29314 Eurydamas     & (1994 CR18)    &   MPC                        & 1J-1       & 5.201    &   L          & trojan        \\
 100624              & (1997 TR28)    &   MPC                        & 1J-1       & 5.201    &   L          & trojan        \\
 22 Kalliope         & (A852 WA)      &   \cite{NesvMorb1998CMDA}    & 4J-4S-1    & 2.906    &  T   & main belt     \\ 
 93 Minerva          & (A867 QA)      &   **                         & 3E+13      & 2.657    &  T   & main belt     \\
 216 Kleopatra       & (A880 GB)      &   **                         & 2M-5J-3    & 2.795    &  T   & main belt     \\
 1803 Zwicky         & (1967 CA)      &   **                         & 4E+15      & 2.413    &  T   & main belt     \\
 1803 Zwicky         & (1967 CA)      &   **                         & 5E-18      & 2.348    &  T   & main belt     \\
 1803 Zwicky         & (1967 CA)      &   **                         & 9M-17      & 2.328    &  T   & main belt     \\
 2171 Kiev           & (1973 QD1)     &   \cite{NesvMorb1998CMDA}    & 7J-2       & 2.256    &  T   & main belt     \\
 2623 Zech           & (A919 SA)      &   **                         & 5M-9       & 2.254    &  T   & main belt     \\
 26420               & (1999 XL103)   &   **                         & 9M-16      & 1.038    &  T   & main belt     \\
 3187 Dalian         & (1977 TO3)     &   **                         & 10M-19     & 2.337    &  T   & main belt     \\
 3390 Demanet        & (1984 ES1)     &   **                         & 6M-11      & 1.017    &  T   & main belt     \\
 3657 Ermolova       & (1978 ST6)     &   **                         & 6M-11      & 1.017    &  T   & main belt     \\
 3865 Lindbloom      & (1988 AY4)     &   **                         & 4J-2S-1    & 2.396    &   L          & main belt     \\
 4901 O Briain       & (1988 VJ)      &   **                         & 13J+4      & 2.370    &  T   & main belt     \\
 4296 van Woerkom    & (1935 SA2)     &   **                         & 15J+4      & 2.154    &  T   & main belt     \\
 4492 Debussy        & (1988 SH)      &   **                         & 4E-19      & 2.825    &  T   & main belt     \\
 4951 Iwamoto        & (1990 BM)      &   \cite{NesvMorb1998CMDA}    & 7J-2       & 2.256    &  T   & main belt     \\
 5319 Petrovskaya    & (1985 RK6)     &   \cite{NesvMorb1998CMDA}    & 7J-2       & 2.256    &  T   & main belt     \\
 5536 Honeycutt      & (1955 QN)      &   **                         & 15J+4      & 2.154    &  T   & main belt     \\
 5872 Sugano         & (1989 SL)      &   **                         & 15J+4      & 2.154    &  T   & main belt     \\
 6615 Plutarchos     & (9512 P-L)     &   **                         & 1V-1E-2    & 2.170    &  T   & main belt     \\
 12218 Fleischer     & (1982 RK)      &   **                         & 4M-7       & 2.212    &  T   & main belt     \\
 15107 Toepperwein   & (2000 CR49)    &   **                         & 7M+12      & 1.063    &  T   & main belt     \\
 15430               & (1998 UR31)    &   **                         & 7J+2       & 2.256    &  T   & main belt     \\
 16525 Shumarinaiko  & (1991 CU2)     &   **                         & 4J-2S-1    & 2.396    &   L          & main belt     \\
 18503               & (1996 PY4)     &   **                         & 13J+4      & 2.370    &  T   & main belt     \\
 21149 Kenmitchell   & (1993 HY5)     &   **                         & 7M-10      & 1.932    &  T   & main belt     \\
 57027               & (2000 UB59)    &   \cite{Szabo2020}           & 3J-2 (Hilda)      & 3.969    &   L          & main belt     \\
 57202               & (2001 QJ53)    &   **                         & 4M-2J-7    & 2.337    &  T   & main belt     \\
 78085               & (2002 LV23)    &   **                         & 5J-4S-1    & 2.303    &  T   & main belt     \\
 78085               & (2002 LV23)    &   **                         & 2E-7       & 2.305    &  T   & main belt     \\
 118303              & (1998 UG)      &   **                         & 3M-5J-4    & 2.264    &  T   & main belt     \\
 26308               & (1998 SM165)   &   Boulder                    & 1N-2       & 47.811   &   L          & TNO           \\ 
 38628 Huya          & (2000 EB173)   &   Boulder                    & 2N-3       & 39.467   &   L          & TNO           \\
 47171 Lempo         & (1999 TC36)    &   Boulder                    & 2N-3       & 39.467   &   L          & TNO           \\
 65489 Ceto          & (2003 FX128)   &   **                         & 1N-6       & 99.452   &  T   & TNO           \\
 60621               & (2000 FE8)     &   Boulder                    & 2N-5       & 55.480   &   L          & TNO           \\
 66652 Borasisi      & (1999 RZ253)   &   **                         & 4N-7       & 43.739   &   LT & TNO           \\
 82075               & (2000 YW134)   &   Boulder                    & 3N-8       & 57.919   &   L          & TNO           \\
 90482 Orcus         & (2004 DW)      &   Boulder                    & 2N-3       & 39.467   &   L          & TNO           \\
 119067              & (2001 KP76)    &   Wikiwand                   & 4N-7       & 43.739   &   L          & TNO           \\
 119979              & (2002 WC19)    &   Boulder                    & 1N-2       & 47.811   &   L          & TNO           \\
 136108 Haumea       & (2003 EL61)    &   \cite{Alonso2009}          & 7N-12      & 43.142   &   LT  & TNO           \\
 139775              & (2001 QG298)   &   Boulder                    & 2N-3       & 39.467   &   L          & TNO           \\
 174567 Varda        & (2003 MW12)    &   Wiki                       & 8N-15      & 45.798   &   LT & TNO           \\ 
 208996              & (2003 AZ84)    &   Boulder                    & 2N-3       & 39.467   &   L          & TNO           \\
 225088 Gonggong     & (2007 OR10)    &   Boulder                    & 3N-10      & 67.209   &   LT & TNO           \\
 341520 Mors-Somnus  & (2007 TY430)   &   Boulder                    & 2N-3       & 39.467   &   L          & TNO           \\ 
 385446 Manwe        & (2003 QW111)   &   Boulder                    & 4N-7       & 43.739   &   L          & TNO           \\
 469420              & (2001 XP254)   &   Boulder                    & 3N-5       & 42.339   &   L          & TNO           \\
 469505              & (2003 FE128)   &   Boulder                    & 1N-2       & 47.811   &   L          & TNO           \\
 506121              & (2016 BP81)    &  **                          & 4N-7       & 43.739   &   LT & TNO           \\
 523624              & (2008 CT190)   &  Boulder                     & 3N-7       & 52.986   &   LT & TNO           \\
 523764              & (2014 WC510)   &  Boulder                     & 2N-3       & 39.467   &   L          & TNO           \\
 525816              & (2005 SF278)   &  Boulder                     & 4N-7       & 43.739   &   L          & TNO           \\ 
 554099              & (2012 KU50)    &  JA                          & 5N-8       & 41.202   &   L          & TNO           \\
 612088              & (1999 CM158)   &  Boulder                     & 2N-3       & 39.467   &   L          & TNO           \\
 612176              & (2000 QL251)   &  Boulder                     & 1N-2       & 47.811   &   L          & TNO           \\
\end{longtable}
}

\begin{table}[H]
\caption{\small {The same as Table~\ref{table2} for the Gaia binary candidates from \cite{Liberato2024}. The 38 objects marked with ** are not found in the previous literature as resonant asteroids.}}

\vskip3mm

\scriptsize
\centerline{\begin{tabular}{llclccc}
\hline
\textbf{Name}         &  \textbf{Prov. name} & \textbf{ Reference  }            & \textbf{MMR}   & \textbf{$a_{MMR}$} & \textbf{Status} & \textbf{class} \\
\hline
    53 Kalypso           &   (A858 GA)     &  \cite{NesvMorb1998CMDA} & 6J-1S-2       & 2.618   &  T    & Main Belt     \\
    217 Eudora           &   (A880 QA)     &     **                   & 2M+5S-6       & 2.872   &  T    & Main Belt     \\
    605 Juvisia          &  (A906QJ)       &     **                   & 7M-20         & 3.067   &  T    & Main Belt     \\
    625 Xenia            &   (A907 CG)     &     **                   &  1V-7         & 2.646   &  T    & Main Belt     \\
    674  Rachele         &   (A908 UK)     &     **                   & 1E-5          & 2.924   &  T    & Main Belt     \\
    923  Herluga         &   (A919 SK)     & \cite{NesvMorb1998CMDA}  & 2J+2S-1       & 2.616   &  T    & Main Belt     \\
    1438  Wendeline      &   (1937 TC)     &    **                    & 5J-2S-2       & 3.173   &  L    & Main Belt     \\
    1963  Bezovec        &   (1975 CB)     &    **                    & 9M-19         & 2.507   &  T    & Main Belt     \\
    3561  Devine         &   (1983 HO)     & \cite{Broz2008}          & 3J-2 (Hilda)  & 3.969   &  L    & Main Belt     \\
    4031 Mueller         &   (1985 CL)     &   **                     & 7E-19         & 1.945   &  T    & Main Belt     \\
    5044 Shestaka        &   (1977 QH4)    &   **                     &  4E-13        & 2.194   &  T    & Main Belt     \\
    5747 Williamina      &   (1991 CO3)    &   **                     & 4J-2S-1       & 2.396   &  L    & Main Belt     \\
    6364 Casarini        &   (1981 ET)     &   **                     & 2E-5M+3       & 2.748   &  T    & Main Belt     \\
    6364 Casarini        &   (1981 ET)     &   **                     & 2M+1S-5       & 2.867   &  T    & Main Belt     \\
    6612 Hachioji        &   (1994 EM1)    &   **                     & 1V-1S-6       & 2.422   &  L    & Main Belt     \\
    7071                 &   (1995 BH4)    &   **                     & 5J-2S-2       & 3.173   &  L    & Main Belt      \\
    8284 Cranach         &   (1991 TT13)   &   **                     & 1V+9          & 3.129   &  T    & Main Belt      \\
    8632 Egleston        &   (1981 FR)     &   **                     & 4J-1S-1       & 2.214   &  T    & Main Belt     \\
    9661 Hohmann         &   (1996 FU13)   &  \cite{Broz2008}         & 3J-2 (Hilda)  & 3.969   &  L    & Main Belt     \\
    11218                &   (1999 JD20)   &   **                     & 9J-4          & 3.029   &  T    & Main Belt     \\
    12914                &   (1998 SJ141)  &   **                     & 3M-7          & 2.680   &  T    & Main Belt     \\
    13840 Wayneanderson  &   (1999 XW31)   &   **                     & 1M-2          & 2.418   &  T    & Main Belt     \\
    14717                &   (2000 CJ82)   &   **                     & 5J-2S-2       & 3.173   &  T    & Main Belt      \\
    15063                &   (1999 AQ3)    &   **                     & 2J+1S-1       & 2.900   &  T    & Main Belt     \\
    15373                &   (1996 WV1)    & \cite{Broz2008}          & 3J-2 (Hilda)  & 3.969   &  L    & Main Belt      \\
    18840 Yoshioba       &   (1999 PT4)    &   **                     & 5E-18         & 2.348   &  T    & Main Belt     \\
    22150                &   (2000 WM49)   &   **                     & 4J-2S-1       & 2.396   &  L    & Main Belt      \\
    31359                &   (1998 UA28)   &   **                     & 7M+12         & 2.182   &  T    & Main Belt      \\
    43326                &   (2000 KH73)   &   **                     & 7M+12         & 2.182   &  T    & Main Belt      \\
    43341                &   (2000 RK62)   &   **                     & 3E+13         & 2.657   &  T    & Main Belt      \\
    43341                &   (2000 RK62)   &   **                     & 8M-17         & 2.518   &  T    & Main Belt      \\
    47211                &   (1999 TX290)  &   **                     & 3E-10         & 2.231   &  T    & Main Belt      \\
    53245                &   (1999 CH152)  &   **                     & 12J-5         & 2.901   &  T    & Main Belt      \\
    55125                &   (2001 QD173)  &   **                     & 2V-4E+3       & 2.522   &  T    & Main Belt     \\
    61574                &   (2000 QE79)   &   **                     & 1M-2          & 2.418   &  T    & Main Belt     \\
    68304                &   (2001 FO97)   &   **                     & 1V-9          & 3.129   &  T    & Main Belt     \\
    68304                &   (2001 FO97)   &   **                     & 16S-3         & 3.129   &  T    & Main Belt     \\
    72039                &   (2000 XG49)   &   **                     & 6E-19         & 2.156   &  T    & Main Belt      \\ 
    87719                &   (2000 SL45)   &   **                     & 3V-20         & 2.562   &  T    & Main Belt      \\
    88538                &   (2001 QG187)  &   **                     & 2E+9          & 2.725   &  T    & Main Belt      \\
    115321               &   (2003 SK219)  &   **                     & 8J-3          & 2.704   &  L    & Main Belt      \\
    115664               &   (2003 UL142)  &   **                     & 1E-3J-4       & 3.060   &  L    & Main Belt     \\
    121297               &   (1999 RU195)  &   **                     & 13J+6         & 3.106   &  L    & Main Belt     \\  
    160293               &   (2003 DK24)   &   **                     & 2J+2S-1       & 2.616   &  T    & Main Belt      \\
    207591               &   (2006 QH56)   &   **                     & 3J-2 (Hilda)  & 3.969   &  L    & Main Belt      \\
    1867 Deiphobus       &   (1971 EA)     &  \cite{2021PSJ}          & 1J-1          & 5.201   &  L    & Jupiter Trojan \\
    3596 Meriones        &   (1985 VO)     &  \cite{2021PSJ}          & 1J-1          & 5.201   &  L    & Jupiter Trojan  \\
    31344 Agathon        &   (1998 OM12)   &  MPC                     & 1J-1          & 5.201   &  L    & Jupiter Trojan  \\
    35277                &   (1996 RV27)   &  MPC                     & 1J-1          & 5.201   &  L    & Jupiter Trojan  \\
    55563                &   (2002 AW34)   &  MPC                     & 1J-1          & 5.201   &  L    & Jupiter Trojan  \\
    60322                &   (1999 XB257)  &  MPC                     & 1J-1          & 5.201   &  L    & Jupiter Trojan  \\
\hline
\end{tabular}}
\label{table3}
\end{table}

Finally, column 3 of Tables \ref{table2} and \ref{table3} lists the sources in which asteroids are reported to be resonant. In addition to standard references, objects classified as resonant on the Minor Planet Center webpage\footnote{\url{https://www.minorplanetcenter.net/}} are marked with “MPC.” Most resonant TNOs are identified in the Boulder database\footnote{\url{https://www.boulder.swri.edu/~buie/kbo/desclass.html}}, while two asteroids are referenced only on wiki-style websites without cited sources\footnote{The asteroid 119067 (2001 KP76) is listed as resonant on Wikiwand: {\url{https://www.wikiwand.com/fr/articles/S/2008_(119067)_1}}, and 174567 Varda (2003 MW12) on The Solar System Fandom: {\url{https://thesolarsystem.fandom.com/wiki/Varda}}}. Objects  marked  with double asterisk (**) are those we could not find in 
the literature, indicating they are the newly discovered resonant binaries. A total of 82 such cases were found.

\section{Conclusions}
\label{conclusions}

\indent

The results of this study showed that the percentage of binary asteroids in mean motion 
resonances largely depends on their dynamical classes. The highest fraction of approximately 30\% resonant asteroids is observed among the TNO 
binaries, meaning that MMRs offer a protective environment for these systems to survive 
perturbations. The lower fraction of resonant binaries in the main asteroid belt suggests that 
the post-collision processes, often responsible for binary formation in this region, do not favour 
long-term resonance captures. This conclusion is supported by the observed clustering of 
main-belt binaries near weaker MMRs and by their transient nature. 
NEA binaries show a high 
resonance fraction, with most cases involving transient captures. These findings are 
somewhat unexpected, given the dynamically unstable nature of the NEA region, where 
frequent planetary encounters and perturbations disrupt these pairs and can lead to the 
breakup of asteroid systems.

Binary systems have a slightly higher representation in resonances (17.4\%) compared to the 
overall asteroid population (14.1\% from \citeauthor{SmirnovDovgalev2018} 
\citeyear{SmirnovDovgalev2018}).~Moreover, stable 
resonances could facilitate binary formation through low-velocity encounters or angular 
momentum exchange. This suggests that, in addition to the primordial nature of the binary 
TNOs, significantly younger systems could be expected among TNOs and in various 
dynamical classes of small bodies, and may also explain the large percentage of binaries 
observed among Trojans and Hilda populations. 

Future research should aim to extend integration timescales, incorporate additional binary 
catalogues, and explore the influence of individual components within these systems. Binary 
asteroids in resonances remain a topic for further investigation. 

\vskip-2mm
\section{Acknowledgements}
This research was supported by the Ministry of Science, Technological 
Development and Innovation of the Republic of Serbia through contract no. 
451-03-136/2025-03/200002 with the Astronomical Observatory (Belgrade).~The 
integrations with the package \texttt{resonances} were performed by E. Smirnov. Authors thank to M. \' Cuk and V. Beni\v sek for useful opinions on the research of binary asteroids, and to the anonymous reviewer for valuable suggestions.

\bibliographystyle{abbrvnat} 
\bibliography{references.bib}

@article{Gallardo2006a,
  title = {Atlas of the Mean Motion Resonances in the {{Solar System}}},
  author = {Gallardo, T},
  year = {2006},
  month = sep,
  journal = {Icarus},
  volume = {184},
  number = {1},
  pages = {29--38},
  issn = {00191035},
  doi = {10.1016/j.icarus.2006.04.001},
  urldate = {2021-05-07},
  langid = {english}
}

@ARTICLE{Giblin1998Icar,
       author = {{Giblin}, Ian and {Petit}, Jean-Marc and {Farinella}, Paolo},
        title = "{Impact Ejecta Rotational Bursting as a Mechanism for Producing Stable Ida-Dactyl Systems}",
      journal = {Icarus},
         year = 1998,
        month = mar,
       volume = {132},
       number = {1},
        pages = {43-52},
          doi = {10.1006/icar.1997.5864},
       adsurl = {https://ui.adsabs.harvard.edu/abs/1998Icar..132...43G}
}

@ARTICLE{SmirnovShevchenko2013Icar,
       author = {{Smirnov}, Evgeny A. and {Shevchenko}, Ivan I.},
        title = "{Massive identification of asteroids in three-body resonances}",
      journal = {Icarus},
         year = 2013,
        month = jan,
       volume = {222},
       number = {1},
        pages = {220-228},
          doi = {10.1016/j.icarus.2012.10.034},
archivePrefix = {arXiv},
       eprint = {1206.1451},
 primaryClass = {astro-ph.EP},
       adsurl = {https://ui.adsabs.harvard.edu/abs/2013Icar..222..220S}
}

@article{SmirnovDovgalev2018,
  title = {Identification of {{Asteroids}} in {{Two-Body Resonances}}},
  author = {Smirnov, E. A. and Dovgalev, I. S.},
  year = {2018},
  month = jul,
  journal = {Solar System Research},
  volume = {52},
  number = {4},
  pages = {347--354},
  issn = {0038-0946},
  doi = {10.1134/S0038094618040056}
}

@ARTICLE{SmirnovDovPop2017,
       author = {{Smirnov}, Evgeny A. and {Dovgalev}, Ilya S. and {Popova}, Elena A.},
        title = "{Asteroids in three-body mean motion resonances with planets}",
      journal = {Icarus},
         year = 2018,
        month = apr,
       volume = {304},
        pages = {24-30},
          doi = {10.1016/j.icarus.2017.09.032},
       adsurl = {https://ui.adsabs.harvard.edu/abs/2018Icar..304...24S}
}

@ARTICLE{Smirnov2023,
       author = {{Smirnov}, E.~A.},
        title = "{A new python package for identifying celestial bodies trapped in mean-motion resonances}",
      journal = {Astronomy and Computing},
         year = 2023,
        month = apr,
       volume = {43},
          eid = {100707},
        pages = {100707},
          doi = {10.1016/j.ascom.2023.100707},
       adsurl = {https://ui.adsabs.harvard.edu/abs/2023A&C....4300707S}
}

@article{Michel2001Sci,
  title={Collisions and gravitational reaccumulation: forming asteroid families and satellites},
  author = {Michel, P. and Benz, W. and Tanga, P. and Richardson, D. C.},
  year = {2001},
  month = nov,
  journal = {Science},
  volume = {294},
  pages = {1696},
  doi = {10.1126/science.1065189}
}

@article{PravecHarris2007,
  title={Binary asteroid population. 1. Angular momentum content},
  author = {Pravec, P. and Harris, A. W.},
  year = {2007},
  month = sep,
  journal = {Icarus},
  volume = {190},
  pages = {250},
  doi = {10.1016/j.icarus.2007.02.023}
}

@article{Rubincam2000,
  title={Radiative Spin-up and Spin-down of Small Asteroids},
  author = {Rubincam, D. P.},
  year = {2000},
  month = nov,
  journal = {Icarus},
  volume = {148},
  pages = {2},
  doi = {10.1006/icar.2000.6485}
}

@article{Bottke2006,
  title={THE YARKOVSKY AND YORP EFFECTS: Implications for Asteroid Dynamics},
  author = {Bottke, W. F. Jr and Vokrouhlick\' y, D. and Rubincam, D. P. and Nesvorn\' y, D.},
  year = {2006},
  month = may,
  journal = {annurev.earth},
  volume = {34},
  pages = {157},
  doi = {10.1146/annurev.earth.34.031405.125154}
}

@INCOLLECTION{Margot2015,
       author = {{Margot}, J. -L. and {Pravec}, P. and {Taylor}, P. and {Carry}, B. and {Jacobson}, S.},
        title = "{Asteroid Systems: Binaries, Triples, and Pairs}",
    booktitle = {Asteroids IV},
         year = 2015,
       editor = {{Michel}, Patrick and {DeMeo}, Francesca E. and {Bottke}, William F.},
        pages = {355-374},
          doi = {10.2458/azu_uapress_9780816532131-ch019},
       adsurl = {https://ui.adsabs.harvard.edu/abs/2015aste.book..355M}
}

@INPROCEEDINGS{Weidenschilling1989,
       author = {{Weidenschilling}, Stuart J. and {Paolicchi}, Paolo and {Zappala}, Vincenzo},
        title = "{Do asteroids have satellites?}",
    booktitle = {Asteroids II},
         year = 1989,
       editor = {{Binzel}, Richard P. and {Gehrels}, Tom and {Matthews}, Mildred Shapley},
        month = jan,
        pages = {643-658},
       adsurl = {https://ui.adsabs.harvard.edu/abs/1989aste.conf..643W}
}

@article{MarcosMarcos2019,
  title={Dancing with Venus in the shadow of the Earth: a pair of genetically related near-Earth asteroids trapped in a mean-motion resonance},
  author = {de la Fuente Marcos, C. and de la Fuente Marcos, R.},
  year = {2018},
  month = nov,
  journal = {Monthly Notices of the Royal Astronomical Society},
  volume = {483},
  pages = {L37},
  doi = {10.1093/mnrasl/sly214,}
}

@online{Johnstons,
  title     = {Asteroids with Satellites},
  author    = {Johnston, Wm. Robert},
  year      = 2024,
  url       = {https://www.johnstonsarchive.net/astro/asteroidmoons.html},
  urldate   = {2024-10-18}
}

@article{Liberato2024,
  title = {Binary Asteroid Candidates in {{{\emph{Gaia}}}} {{DR3}} Astrometry},
  author = {Liberato, L. and Tanga, P. and Mary, D. and Minker, K. and Carry, B. and Spoto, F. and Bartczak, P. and Sicardy, B. and Oszkiewicz, D. and Desmars, J.},
  year = {2024},
  month = aug,
  journal = {Astronomy \& Astrophysics},
  volume = {688},
  pages = {A50},
  issn = {0004-6361, 1432-0746},
  doi = {10.1051/0004-6361/202349122},
  urldate = {2024-10-18},
  copyright = {https://creativecommons.org/licenses/by/4.0}
}

@ARTICLE{Vavilov2022,
       author = {{Vavilov}, Dmitrii E. and {Carry}, Benoit and {Lagain}, Anthony and {Guimpier}, Anthony and {Conway}, Susan and {Devillepoix}, Hadrien and {Bouley}, Sylvain},
        title = "{Evidence for widely-separated binary asteroids recorded by craters on Mars}",
      journal = {Icarus},
         year = 2022,
        month = sep,
       volume = {383},
          eid = {115045},
        pages = {115045},
          doi = {10.1016/j.icarus.2022.115045},
       adsurl = {https://ui.adsabs.harvard.edu/abs/2022Icar..38315045V}
}

@INCOLLECTION{Cook1971,
       author = {{Cook}, A.~F.},
        title = "{624 Hektor: a Binary Asteroid?}",
    booktitle = {NASA Special Publication},
         year = 1971,
       editor = {{Gehrels}, T.},
       volume = {267},
        pages = {155},
       adsurl = {https://ui.adsabs.harvard.edu/abs/1971NASSP.267..155C}
}

@ARTICLE{DunlapGehrels1969AJ,
       author = {{Dunlap}, J.~L. and {Gehrels}, T.},
        title = "{Minor Planets. III. Lightcurves of a Trojan Asteroid}",
      journal = {Astronomical Journal},
         year = 1969,
        month = aug,
       volume = {74},
        pages = {796},
          doi = {10.1086/110860},
       adsurl = {https://ui.adsabs.harvard.edu/abs/1969AJ.....74..796D}
}

@ARTICLE{Binzel1978,
       author = {{Binzel}, Richard P.},
        title = "{Further Support for Minor Planet Multiplicity}",
      journal = {Minor Planet Bulletin},
         year = 1978,
        month = dec,
       volume = {6},
        pages = {18-19},
       adsurl = {https://ui.adsabs.harvard.edu/abs/1978MPBu....6...18B}
}

@ARTICLE{Tedesco1979Sci,
       author = {{Tedesco}, E.~F.},
        title = "{Binary Asteroids: Evidence for Their Existence from Lightcurves}",
      journal = {Science},
         year = 1979,
        month = mar,
       volume = {203},
       number = {4383},
        pages = {905-907},
          doi = {10.1126/science.203.4383.905},
       adsurl = {https://ui.adsabs.harvard.edu/abs/1979Sci...203..905T}
}

@ARTICLE{Binzel1979Sci,
       author = {{Binzel}, R.~P. and {van Flandern}, T.~C.},
        title = "{Minor Planets: The Discovery of Minor Satellites}",
      journal = {Science},
         year = 1979,
        month = mar,
       volume = {203},
       number = {4383},
        pages = {903-905},
          doi = {10.1126/science.203.4383.903},
       adsurl = {https://ui.adsabs.harvard.edu/abs/1979Sci...203..903B}
}

@ARTICLE{Binzel1985Icar,
       author = {{Binzel}, R.~P.},
        title = "{Is 1220 Crocus a precessing, binary asteroid?}",
      journal = {Icarus},
         year = 1985,
        month = jul,
       volume = {63},
       number = {1},
        pages = {99-108},
          doi = {10.1016/0019-1035(85)90022-3},
       adsurl = {https://ui.adsabs.harvard.edu/abs/1985Icar...63...99B}
}

@INCOLLECTION{vanFlandern1979,
       author = {{van Flandern}, T.~C. and {Tedesco}, E.~F. and {Binzel}, R.~P.},
        title = "{Satellites of asteroids.}",
    booktitle = {Asteroids},
         year = 1979,
       editor = {{Gehrels}, Tom and {Matthews}, Mildred Shapley},
        pages = {443-465},
       adsurl = {https://ui.adsabs.harvard.edu/abs/1979aste.book..443V}
}

@ARTICLE{Radau1901BuAs,
       author = {{Radau}, R.},
        title = "{Revue des publications astronomiques. Astronomische Nachrichten, nos. 3663-3721}",
      journal = {Bulletin Astronomique, Serie I},
         year = 1901,
        month = jan,
       volume = {18},
        pages = {423-444},
       adsurl = {https://ui.adsabs.harvard.edu/abs/1901BuAsI..18..423R}
}

@ARTICLE{Andre901,
       author = {{Andr\' e}, Ch.},
        title = "{Sur le systeme forme par la planete double 433 Eros}",
      journal = {Astronomische Nachrichten},
         year = 1901,
        month = jan,
       volume = {155},
        pages = {27},
       adsurl = {}
}

@ARTICLE{Belton1996Icar,
       author = {{Belton}, Michael J.~S. and {Mueller}, Beatrice E.~A. and {D'Amario}, Louis A. and {Byrnes}, Dennis V. and {Klaasen}, Kenneth P. and {Synnott}, Steven and {Breneman}, Herbert and {Johnson}, Torrence V. and {Thomas}, Peter C. and {Veverka}, Joseph and {Harch}, Ann P. and {Davies}, Merton E. and {Merline}, William J. and {Chapman}, Clark R. and {Davis}, Donald and {Denk}, Tilmann and {Neukum}, Gerhard and {Petit}, Jean-Marc and {Greenberg}, Richard and {Storrs}, Alex and {Zellner}, Benjamin},
        title = "{The Discovery and Orbit of 1993 (243)1 Dactyl}",
      journal = {Icarus},
         year = 1996,
        month = mar,
       volume = {120},
       number = {1},
        pages = {185-199},
          doi = {10.1006/icar.1996.0044},
       adsurl = {https://ui.adsabs.harvard.edu/abs/1996Icar..120..185B}
}

@ARTICLE{Mason1994,
       author = {{Mason}, J.~W.},
        title = "{Ida's new moon}",
      journal = {Journal of the British Astronomical Association},
         year = 1994,
        month = jun,
       volume = {104},
        pages = {108-108},
       adsurl = {https://ui.adsabs.harvard.edu/abs/1994JBAA..104..108M}
}

@ARTICLE{Noll2023,
       author = {{Noll}, Keith S. and {Brown}, Michael E. and {Buie}, Marc W. and {Grundy}, William M. and {Levison}, Harold F. and {Marchi}, Simone and {Olkin}, Catherine B. and {Stern}, S. Alan and {Weaver}, Harold A.},
        title = "{Trojan Asteroid Satellites, Rings, and Activity}",
      journal = {Space Science Reviews},
         year = 2023,
        month = oct,
       volume = {219},
       number = {7},
          eid = {59},
        pages = {59},
          doi = {10.1007/s11214-023-01001-w},
       adsurl = {https://ui.adsabs.harvard.edu/abs/2023SSRv..219...59N}
}

@ARTICLE{Nesvorny2021PSJ,
       author = {Nesvorn\'{y}, David and {Li}, Rixin and {Simon}, Jacob B. and {Youdin}, Andrew N. and {Richardson}, Derek C. and {Marschall}, Raphael and {Grundy}, William M.},
        title = "{Binary Planetesimal Formation from Gravitationally Collapsing Pebble Clouds}",
      journal = {The Planetary Science Journal},
         year = 2021,
        month = feb,
       volume = {2},
       number = {1},
          eid = {27},
        pages = {27},
          doi = {10.3847/PSJ/abd858},
archivePrefix = {arXiv},
       eprint = {2011.07042},
 primaryClass = {astro-ph.EP},
       adsurl = {https://ui.adsabs.harvard.edu/abs/2021PSJ.....2...27N}
}

@ARTICLE{NYR2010AJ,
       author = {{Nesvorn{\'y}}, David and {Youdin}, Andrew N. and {Richardson}, Derek C.},
        title = "{Formation of Kuiper Belt Binaries by Gravitational Collapse}",
      journal = {Astronomical Journal},
         year = 2010,
        month = sep,
       volume = {140},
       number = {3},
        pages = {785-793},
          doi = {10.1088/0004-6256/140/3/785},
archivePrefix = {arXiv},
       eprint = {1007.1465},
 primaryClass = {astro-ph.EP},
       adsurl = {https://ui.adsabs.harvard.edu/abs/2010AJ....140..785N}
}

@ARTICLE{Weidenschilling2002,
       author = {{Weidenschilling}, S.~J.},
        title = "{On the Origin of Binary Transneptunian Objects}",
      journal = {Icarus},
         year = 2002,
        month = nov,
       volume = {160},
       number = {1},
        pages = {212-215},
          doi = {10.1006/icar.2002.6952},
       adsurl = {https://ui.adsabs.harvard.edu/abs/2002Icar..160..212W}
}

@ARTICLE{Goldreich2002Natur,
       author = {{Goldreich}, Peter and {Lithwick}, Yoram and {Sari}, Re'em},
        title = "{Formation of Kuiper-belt binaries by dynamical friction and three-body encounters}",
      journal = {Nature},
         year = 2002,
        month = dec,
       volume = {420},
       number = {6916},
        pages = {643-646},
          doi = {10.1038/nature01227},
archivePrefix = {arXiv},
       eprint = {astro-ph/0208490},
 primaryClass = {astro-ph},
       adsurl = {https://ui.adsabs.harvard.edu/abs/2002Natur.420..643G}
}

@ARTICLE{Robinson2020AA,
       author = {{Robinson}, J.~E. and {Fraser}, W.~C. and {Fitzsimmons}, A. and {Lacerda}, P.},
        title = "{Investigating gravitational collapse of a pebble cloud to form transneptunian binaries}",
      journal = {Astronomy \& Astrophysics},
         year = 2020,
        month = nov,
       volume = {643},
          eid = {A55},
        pages = {A55},
          doi = {10.1051/0004-6361/202037456},
archivePrefix = {arXiv},
       eprint = {2008.04207},
 primaryClass = {astro-ph.EP},
       adsurl = {https://ui.adsabs.harvard.edu/abs/2020A&A...643A..55R}
}

@ARTICLE{Fraser2017NatAs,
       author = {{Fraser}, Wesley C. and {Bannister}, Michele T. and {Pike}, Rosemary E. and {Marsset}, Michael and {Schwamb}, Megan E. and {Kavelaars}, J.~J. and {Lacerda}, Pedro and {Nesvorn{\'y}}, David and {Volk}, Kathryn and {Delsanti}, Audrey and {Benecchi}, Susan and {Lehner}, Matthew J. and {Noll}, Keith and {Gladman}, Brett and {Petit}, Jean-Marc and {Gwyn}, Stephen and {Chen}, Ying-Tung and {Wang}, Shiang-Yu and {Alexandersen}, Mike and {Burdullis}, Todd and {Sheppard}, Scott and {Trujillo}, Chad},
        title = "{All planetesimals born near the Kuiper belt formed as binaries}",
      journal = {Nature Astronomy},
         year = 2017,
        month = apr,
       volume = {1},
          eid = {0088},
        pages = {0088},
          doi = {10.1038/s41550-017-0088},
archivePrefix = {arXiv},
       eprint = {1705.00683},
 primaryClass = {astro-ph.EP},
       adsurl = {https://ui.adsabs.harvard.edu/abs/2017NatAs...1E..88F}
}

@INCOLLECTION{Brunini2020,
       author = {{Brunini}, Adrian},
        title = "{Trans-Neptunian binary formation and evolution}",
    booktitle = {The Trans-Neptunian Solar System},
         year = 2020,
       editor = {{Prialnik}, Dina and {Barucci}, Maria Antoinetta and {Young}, Leslie},
        pages = {225-247},
          doi = {10.1016/B978-0-12-816490-7.00010-2},
       adsurl = {https://ui.adsabs.harvard.edu/abs/2020tnss.book..225B}
}

@ARTICLE{Lopez2021MNRAS,
       author = {{L{\'o}pez}, Mar{\'\i}a C. and {Brunini}, A.},
        title = "{Binary-binary close encounters in the Kuiper Belt}",
      journal = {Monthly Notices of the Royal Astronomical Society},
         year = 2021,
        month = jul,
       volume = {505},
       number = {1},
        pages = {236-244},
          doi = {10.1093/mnras/stab1250},
       adsurl = {https://ui.adsabs.harvard.edu/abs/2021MNRAS.505..236L}
}

@ARTICLE{Lawler2024,
       author = {{Lawler}, Samantha M. and {Pike}, Rosemary E.},
        title = "{Small Bodies in the Distant Solar System}",
      journal = {arXiv e-prints},
         year = 2024,
        month = oct,
          eid = {arXiv:2410.04338},
        pages = {arXiv:2410.04338},
          doi = {10.48550/arXiv.2410.04338},
archivePrefix = {arXiv},
       eprint = {2410.04338},
          URL = {  },
 primaryClass = {astro-ph.EP},
       adsurl = {https://ui.adsabs.harvard.edu/abs/2024arXiv241004338L}
}

@ARTICLE{Durda2004Icar,
       author = {{Durda}, Daniel D. and {Bottke}, William F. and {Enke}, Brian L. and {Merline}, William J. and {Asphaug}, Erik and {Richardson}, Derek C. and {Leinhardt}, Zo{\"e} M.},
        title = "{The formation of asteroid satellites in large impacts: results from numerical simulations}",
      journal = {Icarus},
         year = 2004,
        month = jul,
       volume = {170},
       number = {1},
        pages = {243-257},
          doi = {10.1016/j.icarus.2004.04.003},
       adsurl = {https://ui.adsabs.harvard.edu/abs/2004Icar..170..243D}
}

@ARTICLE{Durda1996Icar,
       author = {{Durda}, Daniel D.},
        title = "{The Formation of Asteroidal Satellites in Catastrophic Collisions}",
      journal = {Icarus},
         year = 1996,
        month = mar,
       volume = {120},
       number = {1},
        pages = {212-219},
          doi = {10.1006/icar.1996.0046},
       adsurl = {https://ui.adsabs.harvard.edu/abs/1996Icar..120..212D}
}

@ARTICLE{Doressoundiram1997,
       author = {{Doressoundiram}, A. and {Paolicchi}, P. and {Verlicchi}, A. and {Cellino}, A.},
        title = "{The formation of binary asteroids as outcomes of catastrophic collisions}",
      journal = {Planetary and Space Science},
         year = 1997,
        month = jul,
       volume = {45},
       number = {7},
        pages = {757-770},
          doi = {10.1016/S0032-0633(97)00037-8},
       adsurl = {https://ui.adsabs.harvard.edu/abs/1997P&SS...45..757D}
}

@ARTICLE{Wimarsson2024Icar,
       author = {{Wimarsson}, John and {Xiang}, Zhen and {Ferrari}, Fabio and {Jutzi}, Martin and {Madeira}, Gustavo and {Raducan}, Sabina D. and {S{\'a}nchez}, Paul},
        title = "{Rapid formation of binary asteroid systems post rotational failure: A recipe for making atypically shaped satellites}",
      journal = {Icarus},
         year = 2024,
        month = oct,
       volume = {421},
          eid = {116223},
        pages = {116223},
          doi = {10.1016/j.icarus.2024.116223},
archivePrefix = {arXiv},
       eprint = {2407.15543},
 primaryClass = {astro-ph.EP},
       adsurl = {https://ui.adsabs.harvard.edu/abs/2024Icar..42116223W}
}

@ARTICLE{Cuk2007ApJ,
       author = {{{\'C}uk}, Matija},
        title = "{Formation and Destruction of Small Binary Asteroids}",
      journal = {The Astrophysical Journal Letters},
         year = 2007,
        month = apr,
       volume = {659},
       number = {1},
        pages = {L57-L60},
          doi = {10.1086/516572},
       adsurl = {https://ui.adsabs.harvard.edu/abs/2007ApJ...659L..57C}
}

@ARTICLE{WalshRichardson2008,
       author = {{Walsh}, Kevin J. and {Richardson}, Derek C.},
        title = "{A steady-state model of NEA binaries formed by tidal disruption of gravitational aggregates}",
      journal = {Icarus},
         year = 2008,
        month = feb,
       volume = {193},
       number = {2},
        pages = {553-566},
          doi = {10.1016/j.icarus.2007.08.020},
       adsurl = {https://ui.adsabs.harvard.edu/abs/2008Icar..193..553W}
}

@ARTICLE{VokrouCapek2002Icar,
       author = {{Vokrouhlick{\'y}}, D. and {{\v{C}}apek}, D.},
        title = "{YORP-Induced Long-Term Evolution of the Spin State of Small Asteroids and Meteoroids: Rubincam's Approximation}",
      journal = {Icarus},
         year = 2002,
        month = oct,
       volume = {159},
       number = {2},
        pages = {449-467},
          doi = {10.1006/icar.2002.6918},
       adsurl = {https://ui.adsabs.harvard.edu/abs/2002Icar..159..449V}
}

@ARTICLE{JacobsonScheeres2011Icar,
       author = {{Jacobson}, Seth A. and {Scheeres}, Daniel J.},
        title = "{Dynamics of rotationally fissioned asteroids: Source of observed small asteroid systems}",
      journal = {Icarus},
         year = 2011,
        month = jul,
       volume = {214},
       number = {1},
        pages = {161-178},
          doi = {10.1016/j.icarus.2011.04.009},
archivePrefix = {arXiv},
       eprint = {1404.0801},
 primaryClass = {astro-ph.EP},
       adsurl = {https://ui.adsabs.harvard.edu/abs/2011Icar..214..161J}
}

@ARTICLE{Scheeres2009CeMDA,
       author = {{Scheeres}, D.~J.},
        title = "{Stability of the planar full 2-body problem}",
      journal = {Celestial Mechanics and Dynamical Astronomy},
         year = 2009,
        month = jun,
       volume = {104},
       number = {1-2},
        pages = {103-128},
          doi = {10.1007/s10569-009-9184-7},
       adsurl = {https://ui.adsabs.harvard.edu/abs/2009CeMDA.104..103S}
}

@INCOLLECTION{Noll2008book,
       author = {{Noll}, K.~S. and {Grundy}, W.~M. and {Chiang}, E.~I. and {Margot}, J. -L. and {Kern}, S.~D.},
        title = "{Binaries in the Kuiper Belt}",
    booktitle = {The Solar System Beyond Neptune},
         year = 2008,
       editor = {{Barucci}, M.~A. and {Boehnhardt}, H. and {Cruikshank}, D.~P. and {Morbidelli}, A. and {Dotson}, Renee},
        pages = {345-363},
          doi = {10.48550/arXiv.astro-ph/0703134},
       adsurl = {https://ui.adsabs.harvard.edu/abs/2008ssbn.book..345N}
}

@ARTICLE{Scheeres2007Icar,
       author = {{Scheeres}, D.~J.},
        title = "{Rotational fission of contact binary asteroids}",
      journal = {Icarus},
         year = 2007,
        month = aug,
       volume = {189},
       number = {2},
        pages = {370-385},
          doi = {10.1016/j.icarus.2007.02.015},
       adsurl = {https://ui.adsabs.harvard.edu/abs/2007Icar..189..370S}
}

@ARTICLE{WalshRichar2006Icar,
       author = {{Walsh}, Kevin J. and {Richardson}, Derek C.},
        title = "{Binary near-Earth asteroid formation: Rubble pile model of tidal disruptions}",
      journal = {Icarus},
         year = 2006,
        month = jan,
       volume = {180},
       number = {1},
        pages = {201-216},
          doi = {10.1016/j.icarus.2005.08.015},
       adsurl = {https://ui.adsabs.harvard.edu/abs/2006Icar..180..201W}
}

@ARTICLE{WalshRM2008Nature,
       author = {{Walsh}, Kevin J. and {Richardson}, Derek C. and {Michel}, Patrick},
        title = "{Rotational breakup as the origin of small binary asteroids}",
      journal = {Nature},
         year = 2008,
        month = jul,
       volume = {454},
       number = {7201},
        pages = {188-191},
          doi = {10.1038/nature07078},
       adsurl = {https://ui.adsabs.harvard.edu/abs/2008Natur.454..188W}
}

@ARTICLE{MeloshStansberryIcar1991,
       author = {{Melosh}, H.~J. and {Stansberry}, J.~A.},
        title = "{Doublet craters and the tidal disruption of binary asteroids}",
      journal = {Icarus},
         year = 1991,
        month = nov,
       volume = {94},
       number = {1},
        pages = {171-179},
          doi = {10.1016/0019-1035(91)90148-M},
       adsurl = {https://ui.adsabs.harvard.edu/abs/1991Icar...94..171M}
}

@ARTICLE{Cook2003Icar,
       author = {{Cook}, Cheryl M. and {Melosh}, H. Jay and {Bottke}, William F.},
        title = "{Doublet craters on Venus}",
      journal = {Icarus},
         year = 2003,
        month = sep,
       volume = {165},
       number = {1},
        pages = {90-100},
          doi = {10.1016/S0019-1035(03)00177-5},
       adsurl = {https://ui.adsabs.harvard.edu/abs/2003Icar..165...90C}
}

@ARTICLE{Miljkovic2013EPSL,
       author = {{Miljkovi{\'c}}, Katarina and {Collins}, Gareth S. and {Mannick}, Sahil and {Bland}, Philip A.},
        title = "{Morphology and population of binary asteroid impact craters}",
      journal = {Earth and Planetary Science Letters},
         year = 2013,
        month = feb,
       volume = {363},
        pages = {121-132},
          doi = {10.1016/j.epsl.2012.12.033},
       adsurl = {https://ui.adsabs.harvard.edu/abs/2013E&PSL.363..121M}
}

@ARTICLE{Rosaev2024CeMDA,
       author = {{Rosaev}, Alexey},
        title = "{Resonance perturbation of (5026) Martes and 2005 WW113 asteroid pair}",
      journal = {Celestial Mechanics and Dynamical Astronomy},
         year = 2024,
        month = oct,
       volume = {136},
       number = {5},
          eid = {41},
        pages = {41},
          doi = {10.1007/s10569-024-10209-z},
       adsurl = {https://ui.adsabs.harvard.edu/abs/2024CeMDA.136...41R}
}

@ARTICLE{Borisov2024CoSka,
       author = {{Borisov}, G. and {Todorovi{\'c}}, N. and {Vchkova-Bebekovska}, E. and {Kostov}, A. and {Apostolovska}, G.},
        title = "{The possible dual nature of the asteroid (12499) 1998 FR47}",
      journal = {Contributions of the Astronomical Observatory Skalnate Pleso},
         year = 2024,
        month = nov,
       volume = {54},
       number = {4},
        pages = {57-77},
          doi = {10.31577/caosp.2024.54.4.57},
archivePrefix = {arXiv},
       eprint = {2411.11994},
 primaryClass = {astro-ph.EP},
       adsurl = {https://ui.adsabs.harvard.edu/abs/2024CoSka..54d..57B}
}

@ARTICLE{Pravec2019Icar,
       author = {{Pravec}, P. and {Fatka}, P. and {Vokrouhlick{\'y}}, D. and {Scheirich}, P. and {{\v{D}}urech}, J. and {Scheeres}, D.~J. and {Ku{\v{s}}nir{\'a}k}, P. and {Hornoch}, K. and {Gal{\'a}d}, A. and {Pray}, D.~P. and {Krugly}, Yu. N. and {Burkhonov}, O. and {Ehgamberdiev}, Sh. A. and {Pollock}, J. and {Moskovitz}, N. and {Thirouin}, A. and {Ortiz}, J.~L. and {Morales}, N. and {Hus{\'a}rik}, M. and {Inasaridze}, R. Ya. and {Oey}, J. and {Polishook}, D. and {Hanu{\v{s}}}, J. and {Ku{\v{c}}{\'a}kov{\'a}}, H. and {Vra{\v{s}}til}, J. and {Vil{\'a}gi}, J. and {Gajdo{\v{s}}}, {\v{S}}. and {Korno{\v{s}}}, L. and {Vere{\v{s}}}, P. and {Gaftonyuk}, N.~M. and {Hromakina}, T. and {Sergeyev}, A.~V. and {Slyusarev}, I.~G. and {Ayvazian}, V.~R. and {Cooney}, W.~R. and {Gross}, J. and {Terrell}, D. and {Colas}, F. and {Vachier}, F. and {Slivan}, S. and {Skiff}, B. and {Marchis}, F. and {Ergashev}, K.~E. and {Kim}, D. -H. and {Aznar}, A. and {Serra-Ricart}, M. and {Behrend}, R. and {Roy}, R. and {Manzini}, F. and {Molotov}, I.~E.},
        title = "{Asteroid pairs: A complex picture}",
      journal = {Icarus},
         year = 2019,
        month = nov,
       volume = {333},
        pages = {429-463},
          doi = {10.1016/j.icarus.2019.05.014},
archivePrefix = {arXiv},
       eprint = {1901.05492},
 primaryClass = {astro-ph.EP},
       adsurl = {https://ui.adsabs.harvard.edu/abs/2019Icar..333..429P}
}

@ARTICLE{Duddy2012AA,
       author = {{Duddy}, S.~R. and {Lowry}, S.~C. and {Wolters}, S.~D. and {Christou}, A. and {Weissman}, P. and {Green}, S.~F. and {Rozitis}, B.},
        title = "{Physical and dynamical characterisation of the unbound asteroid pair 7343-154634}",
      journal = {Astronomy \& Astrophysics},
         year = 2012,
        month = mar,
       volume = {539},
          eid = {A36},
        pages = {A36},
          doi = {10.1051/0004-6361/201118302},
       adsurl = {https://ui.adsabs.harvard.edu/abs/2012A&A...539A..36D}
}

@ARTICLE{PravecVokrouhlicky2009Icar,
       author = {{Pravec}, P. and {Vokrouhlick{\'y}}, D.},
        title = "{Significance analysis of asteroid pairs}",
      journal = {Icarus},
         year = 2009,
        month = dec,
       volume = {204},
       number = {2},
        pages = {580-588},
          doi = {10.1016/j.icarus.2009.07.004},
       adsurl = {https://ui.adsabs.harvard.edu/abs/2009Icar..204..580P}
}

@ARTICLE{Sonnett2015ApJ,
       author = {{Sonnett}, S. and {Mainzer}, A. and {Grav}, T. and {Masiero}, J. and {Bauer}, J.},
        title = "{Binary Candidates in the Jovian Trojan and Hilda Populations from NEOWISE Light Curves}",
      journal = {Astrophysical Journal},
         year = 2015,
        month = feb,
       volume = {799},
       number = {2},
          eid = {191},
        pages = {191},
          doi = {10.1088/0004-637X/799/2/191},
archivePrefix = {arXiv},
       eprint = {1412.1853},
 primaryClass = {astro-ph.EP},
       adsurl = {https://ui.adsabs.harvard.edu/abs/2015ApJ...799..191S}
}

@ARTICLE{Compere2013AA,
       author = {{Comp{\`e}re}, A. and {Farrelly}, D. and {Lema{\^\i}tre}, A. and {Hestroffer}, D.},
        title = "{A possible mechanism to explain the lack of binary asteroids among the Plutinos}",
      journal = {Astronomy \& Astrophysics},
         year = 2013,
        month = oct,
       volume = {558},
          eid = {A4},
        pages = {A4},
          doi = {10.1051/0004-6361/201321137},
       adsurl = {https://ui.adsabs.harvard.edu/abs/2013A&A...558A...4C}
}

@ARTICLE{Thirouin2018AJ,
       author = {{Thirouin}, Audrey and {Sheppard}, Scott S.},
        title = "{The Plutino Population: An Abundance of Contact Binaries}",
      journal = {     The Astronomical Journal},
         year = 2018,
        month = jun,
       volume = {155},
       number = {6},
          eid = {248},
        pages = {248},
          doi = {10.3847/1538-3881/aac0ff},
archivePrefix = {arXiv},
       eprint = {1804.09695},
 primaryClass = {astro-ph.EP},
       adsurl = {https://ui.adsabs.harvard.edu/abs/2018AJ....155..248T}
}

@ARTICLE{Brunini2023MNRAS,
       author = {{Brunini}, Adri{\'a}n},
        title = "{The origin of an overpopulation of contact binary plutinos}",
      journal = {Monthly Notices of the Royal Astronomical Society},
         year = 2023,
        month = sep,
       volume = {524},
       number = {1},
        pages = {L45-L49},
          doi = {10.1093/mnrasl/slad076},
       adsurl = {https://ui.adsabs.harvard.edu/abs/2023MNRAS.524L..45B}
}

@ARTICLE{Thirouin2024PSJ,
       author = {{Thirouin}, Audrey and {Sheppard}, Scott S.},
        title = "{Rotational Study of 5:3 and 7:4 Resonant Objects within the Main Classical Trans-Neptunian Belt}",
      journal = {The Planetary Science Journal},
         year = 2024,
        month = apr,
       volume = {5},
       number = {4},
          eid = {84},
        pages = {84},
          doi = {10.3847/PSJ/ad2933},
archivePrefix = {arXiv},
       eprint = {2402.05269},
 primaryClass = {astro-ph.EP},
       adsurl = {https://ui.adsabs.harvard.edu/abs/2024PSJ.....5...84T}
}

@ARTICLE{NesvMorb1998CMDA,
       author = {{Nesvorn{\'y}}, D. and {Morbidelli}, A.},
        title = "{Three-Body Mean Motion Resonances and the Chaotic Structure of the Asteroid Belt}",
      journal = {The Astronomical Journal},
         year = 1998,
        month = dec,
       volume = {116},
       number = {6},
        pages = {3029-3037},
          doi = {10.1086/300632},
       adsurl = {https://ui.adsabs.harvard.edu/abs/1998AJ....116.3029N}
}

@article{RNF2002,
    author = {Roig, F. and Nesvorn{\'y}, D. and Ferraz-Mello, S.},
    title = {Asteroids in the 2 : 1 resonance with Jupiter: dynamics and size distribution},
    journal = {Monthly Notices of the Royal Astronomical Society},
    volume = {335},
    number = {2},
    pages = {417-431},
    year = {2002},
    month = {09},
    issn = {0035-8711},
    doi = {10.1046/j.1365-8711.2002.05635.x},
    url = {https://doi.org/10.1046/j.1365-8711.2002.05635.x},
}

@ARTICLE{Szabo2020,
       author = {{Szab{\'o}}, Gyula M. and {Kiss}, Csaba and {Szak{\'a}ts}, R{\'o}bert and {P{\'a}l}, Andr{\'a}s and {Moln{\'a}r}, L{\'a}szl{\'o} and {S{\'a}rneczky}, Kriszti{\'a}n and {Vink{\'o}}, J{\'o}zsef and {Szab{\'o}}, R{\'o}bert and {Marton}, G{\'a}bor and {Kiss}, L{\'a}szl{\'o} L.},
        title = "{Rotational Properties of Hilda Asteroids Observed by the K2 Mission}",
      journal = {The Astrophysical Journal Supplement Series},
         year = 2020,
        month = mar,
       volume = {247},
       number = {1},
          eid = {34},
        pages = {34},
          doi = {10.3847/1538-4365/ab6b23},
archivePrefix = {arXiv},
       eprint = {2001.06656},
 primaryClass = {astro-ph.EP},
       adsurl = {https://ui.adsabs.harvard.edu/abs/2020ApJS..247...34S}
}

@ARTICLE{Alonso2009,
       author = {{Pinilla-Alonso}, N. and {Brunetto}, R. and {Licandro}, J. and {Gil-Hutton}, R. and {Roush}, T.~L. and {Strazzulla}, G.},
        title = "{The surface of (136108) Haumea (2003 EL\{61\}), the largest carbon-depleted object in the trans-Neptunian belt}",
      journal = {Astronomy \& Astrophysics},
         year = 2009,
        month = mar,
       volume = {496},
       number = {2},
        pages = {547-556},
          doi = {10.1051/0004-6361/200809733},
archivePrefix = {arXiv},
       eprint = {0803.1080},
 primaryClass = {astro-ph},
       adsurl = {https://ui.adsabs.harvard.edu/abs/2009A&A...496..547P}
}

@ARTICLE{2021PSJ,
       author = {{McNeill}, A. and {Erasmus}, N. and {Trilling}, D.~E. and {Emery}, J.~P. and {Tonry}, J.~L. and {Denneau}, L. and {Flewelling}, H. and {Heinze}, A. and {Stalder}, B. and {Weiland}, H.~J.},
        title = "{Comparison of the Physical Properties of the L4 and L5 Trojan Asteroids from ATLAS Data}",
      journal = {The Planetary Science Journal},
         year = 2021,
        month = feb,
       volume = {2},
       number = {1},
          eid = {6},
        pages = {6},
          doi = {10.3847/PSJ/abcccd},
archivePrefix = {arXiv},
       eprint = {2101.04602},
 primaryClass = {astro-ph.EP},
       adsurl = {https://ui.adsabs.harvard.edu/abs/2021PSJ.....2....6M}
}

@article{Broz2008,
    author = {Bro{\v z}, M. and Vokrouhlick{\'y}, D.},
    title = {Asteroid families in the first-order resonances with Jupiter},
    journal = {Monthly Notices of the Royal Astronomical Society},
    volume = {390},
    number = {2},
    pages = {715-732},
    year = {2008},
    month = {10},
    issn = {0035-8711},
    doi = {10.1111/j.1365-2966.2008.13764.x},
    url = {https://doi.org/10.1111/j.1365-2966.2008.13764.x},
    eprint = {https://academic.oup.com/mnras/article-pdf/390/2/715/3439700/mnras0390-0715.pdf},
}

@ARTICLE{Forgacs-Dajka2023ApJS,
       author = {{Forg{\'a}cs-Dajka}, E. and {K{\H{o}}v{\'a}ri}, E. and {Kov{\'a}cs}, T. and {Kiss}, Cs. and {S{\'a}ndor}, Zs.},
        title = "{A Dynamical Survey of the Trans-Neptunian Region. I. Mean-motion Resonances with Neptune}",
      journal = {The Astrophysical Journal Supplement Series},
         year = 2023,
        month = may,
       volume = {266},
       number = {1},
          eid = {5},
        pages = {5},
          doi = {10.3847/1538-4365/acc4c8},
archivePrefix = {arXiv},
       eprint = {2302.01221},
 primaryClass = {astro-ph.EP},
       adsurl = {https://ui.adsabs.harvard.edu/abs/2023ApJS..266....5F}
}

\end{document}